\documentclass[fleqn,10pt]{wlscirep}
\usepackage[utf8]{inputenc}
\usepackage[T1]{fontenc}
\title{Theory of the $sp - d$ coupling of transition metal impurities 
with free carriers in ZnO}
\author[1,*]{Anna Ciechan}
\author[1,**]{Piotr Bogus{\l}awski}
\affil[1]{Institute of Physics, Polish Academy of Sciences, al. 
Lotnik\'{o}w 32/46, 02-668 Warsaw, Poland}
\affil[*]{ciechan@ifpan.edu.pl}
\affil[**]{bogus@ifpan.edu.pl}

\begin{abstract} 
The $s,p-d$ exchange coupling between the spins of band carriers and of 
transition metal (TM) dopants ranging from Ti to Cu in ZnO is studied 
within the density functional theory. The $+U$ corrections are included 
to reproduce the experimental ZnO band gap and the dopant levels. The $p-
d$ coupling reveals unexpectedly complex features. In particular, (i) the 
$p-d$ coupling constants $N_0\beta$ vary about 10 times when going from V 
to Cu, (ii) not only the value but also the sign of $N_0\beta$ depends on 
the charge state of the dopant, (iii) the $p-d$ coupling with the heavy 
holes and the light holes is not the same; in the case of Fe, Co and Ni, 
$N_0\beta$s for the two subbands can differ twice, and for Cu the 
opposite sign of the coupling is found for light and heavy holes. The 
main features of the $p-d$ coupling are determined by the $p-d$ 
hybridization between the $d$(TM) and $p$(O) orbitals. In contrast, the 
$s-d$ coupling constant $N_0\alpha$ is almost the same for all TM ions, 
and does not depend on the charge state of the dopant. The TM-induced 
spin polarization of the $p$(O) orbitals contributes to the $s-d$ 
coupling, enhancing $N_0\alpha$. 
\end{abstract}

\begin{document}
\flushbottom
\maketitle
\thispagestyle{empty}

\section{Introduction}
The $s,p-d$ coupling between free carriers and the localized $d$-
electrons of the TM dopants constitutes the basic feature of diluted 
magnetic semiconductors (DMSs).~\cite{book2010, Dietl2014} The interest 
in this class of materials sharply raised after it was demonstrated that 
the $s,p-d$ coupling enables a control of electronic properties by 
magnetic field, and vice versa, control of magnetic properties by 
electric field. Those properties of the DMS-based structures were applied 
to obtain novel spintronic functionalities. 
Next, free carriers mediate magnetic interactions between the TM ions in 
DMSs through the $s,p-d$ coupling, and lead to collective magnetism under 
appropriate conditions. At the basic level, the electron-electron Coulomb 
coupling includes the spin-dependent exchange channel, which can be 
represented in the effective Heisenberg form. In DMSs, the dominant 
"effective" mechanism of coupling with localized spins of magnetic atoms 
is different for electrons and holes because of the different symmetries 
of their wave functions. 

Theory of coupling between the conduction $s$ electrons and the localized 
$4f$ shell in the rare earth metals was elaborated by Liu,~\cite{Liu_ex} 
and assumed to be operative also in the case of the $s-3d$ coupling in 
semiconductors. This mechanism is referred to as direct exchange, because 
it originates in the direct intra-atomic exchange coupling between the 
overlapping 
wave functions of 
$s$ and $d$ (or $f$) electrons. In turn, the mechanism of the $p-d$ 
coupling proposed by Anderson,~\cite{Anderson} $i.e.$, the kinetic 
exchange, relies on the symmetry-allowed hybridization of the $d$(TM) 
shell with the hole states, or, in the real space picture, with the $p$ 
orbitals of the anion neighbors of the TM dopant. Both the Liu's and the 
Anderson's models are widely used to study and explain magnetic 
properties of DMSs.~\cite{Schrieffer, Larson, Bhattacharjee,  
Blinowski2001, Kacman, Beaulac} In parallel, the $s,p-d$ coupling was 
evaluated with the density functional theory (DFT) calculations for III-V 
and II-VI semiconductors.~\cite{Zunger1987, Sanvito, Sandratskii, 
Chanier, Sato} This approach treats all electrons on the same footing and 
includes automatically intra- as well as inter-atomic interactions. 

In this work, we employ the DFT calculations to study the $s,p-d$ 
exchange coupling for the TM impurities ranging from Ti to Cu in ZnO. Our 
study of the entire 3$d$ TM series reveals not only the properties of 
individual dopants, but also trends in the $s-d$ and the $p-d$ couplings, 
assessing their general features. Interpretation of the results based on 
the analysis of the relevant wave functions reveals the role of the Liu's 
and the Anderson's mechanisms, but first of all it shows the dominant 
role played by the $p-d$ hybridization.  This latter effect leads to the 
spin polarization of not only the $p$(O) orbitals of host oxygen ions in 
the vicinity of the TM ion (leading to the spin splitting of the valence 
band maximum (VBM) and finite $N_0\beta$s), but also of their $s$(O) 
orbitals, what provides an additional contribution to the $s-d$ coupling. 
In the literature, the $p-d$ exchange constant $N_0\beta$ is often 
tacitly assumed to be a constant independent of factors such as the 
charge state of the dopant. This picture is not compatible with the 
dominant role of the hybridization, which depends on the inverse energy 
distance between the $d$(TM)-induced levels and the VBM. Indeed, the 
pronounced dependence of the TM level energies on its charge state can be 
reflected not only in the magnitude, but also in the sign of $N_0\beta$. 
Moreover, $N_0\beta$ can be different for the light and heavy hole 
subbands depending on the detailed electronic structure of a TM ion. 

\section{Results}
\subsection{TM impurity levels in ZnO}
\begin{figure}[t!]
\begin{center}
\includegraphics[width=\textwidth]{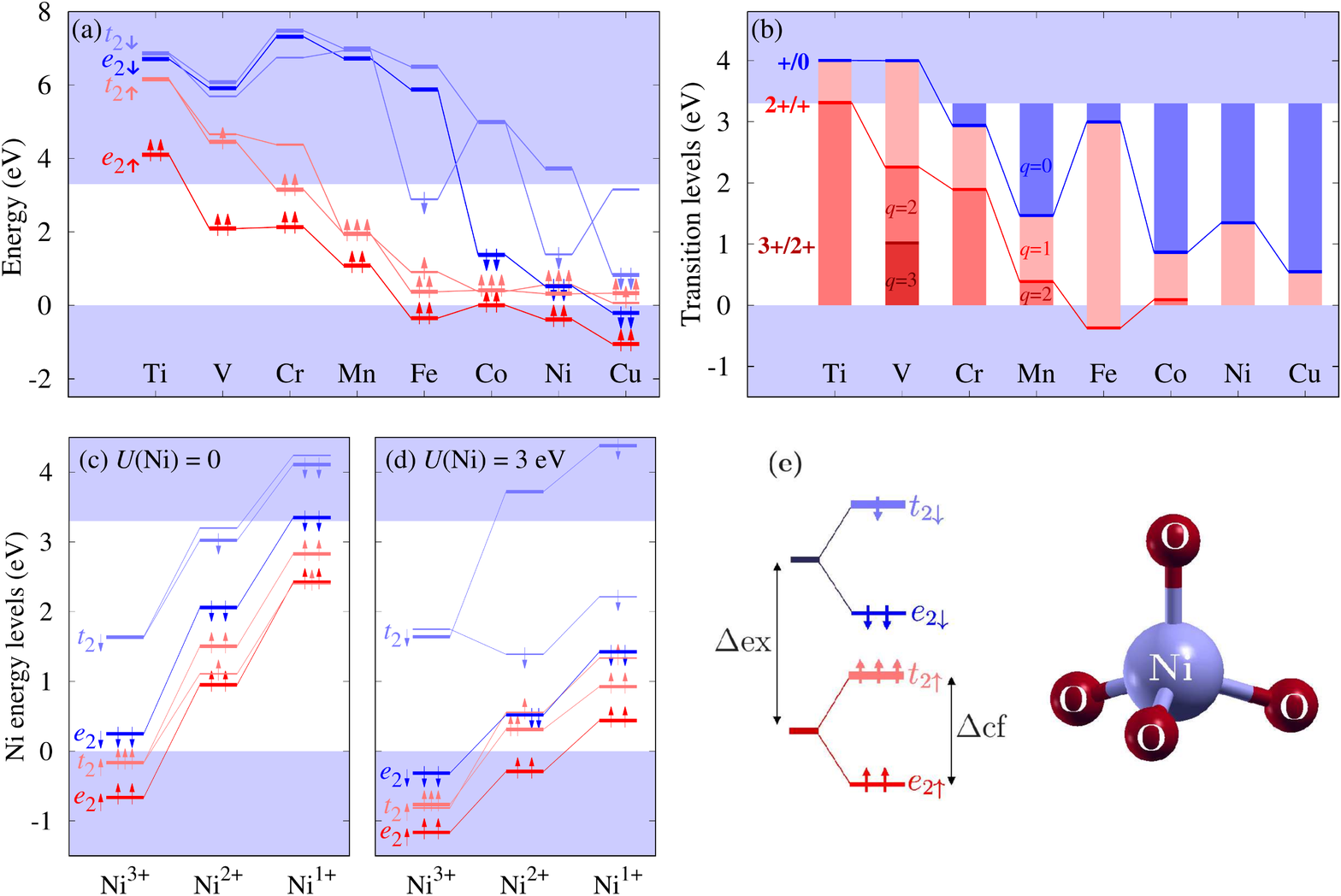}\\
\end{center}
\caption{\label{fig:levels} \small
(a) Single particle levels of 
charge neutral TM ($q=0$) 
and (b) transition levels $\varepsilon(q/q')$ of TMs in ZnO obtained with 
the optimized $U$(TM) values (see Method of calculations). 
As it is schematically indicated, Fe$^{4+}$ is stable in highly $p$-doped 
samples. 
Charge state dependence of the Ni levels for (c) $U$(Ni)=0 and (d) 
$U$(Ni)=3~eV. 
In each panel, the levels of TM below the VBM are shown only 
schematically. 
(e) Schematics of exchange ($\Delta\mathrm{ex}$) and tetrahedral crystal 
field ($\Delta\mathrm{cf}$) splittings of Ni$^{2+}$. In the wurtzite 
structure the three planar bonds are not equivalent with the vertical 
one, which causes further splitting of triplet levels. Arrows denote 
spins. 
}
\end{figure}

Several magnetic properties of a TM dopant are determined by its energy 
levels relative to the VBM and the conduction band minimum (CBM). 
An exemplary band structure of TM-doped ZnO is discussed in 
Supplementary Information, see Figs~\ref{fig:Mn1} and \ref{fig:Mn2}, while
the relevant results, necessary to 
understand the mechanism of the $s,p-d$ coupling, are given in 
Fig.~\ref{fig:levels}. 
We first recall that the $d$(TM) shell of a substitutional TM ion in a 
zinc blende crystal is split into a $e_2$ doublet 
and a $t_2$ triplet higher in energy. Both states are spin split by the 
exchange coupling, stronger than the crystal field, and all TM ions 
in ZnO are in the high spin state, see Fig.~\ref{fig:levels} (e). 
Next, in ZnO the triplets are further split by the uniaxial 
wurtzite crystal field into singlets and doublets. 
This holds also for the $p$(O)-derived VBM, which is split 
by 64 meV into a light hole singlet and a heavy hole doublet, 
denoted by $A_1$ and $E_2$ in the following. 

When the $U$(TM) corrections are employed, the splitting of a $t_2$ state 
depends on its occupation. The $U$-induced contribution $V_U$ to the 
Kohn-Sham potential is~\cite{Cococcioni}
\begin{equation}
V_U|\psi_{k\nu}^\sigma\rangle = U \sum_{m,\sigma} 
(1/2 - 
\lambda_m^\sigma)|\phi_m\rangle\langle\phi_m|\psi_{k\nu}^\sigma\rangle,
\label{eq:U} 
\end{equation}
where $\phi_m$ are the localized $d$ orbitals occupied by 
$\lambda_m^\sigma $ 
electrons, and $\psi_{k\nu}^\sigma$ are the Kohn-Sham states for the 
wave vector $k$, band $\nu$, and spin $\sigma$. 
The $V_U$ potential only acts on the contribution of the $m$th $d$(TM) 
orbital to the given $(\nu,k,\sigma)$ state.  
As a result of the $U$(TM) correction, which favours fully occupied or 
completely empty orbitals over the partially occupied 
ones,~\cite{Cococcioni} 
the splitting of the fully occupied $t_2$ is small, about 0.1~eV, which 
is close to the crystal field splitting of the VBM. 
Otherwise, the splitting is more pronounced, and can exceed the crystal 
field splitting. 
The calculated energy levels of neutral TM$^{2+}$ ions are shown in 
Fig.~\ref{fig:levels} (a). 
Considering the series Fe -- Cu we see that $t_{2\uparrow}$ is very close 
to the VBM. Interestingly, this trend reflects the $d$-shell energies of 
isolated TM atoms, see Supplementary Information, 
Fig.~\ref{fig:TMatoms}. 

Figures~\ref{fig:levels} (c), (d) show the charge state dependence of the 
energies of the  Ni levels as an example. With the increasing occupation 
of the $d$ shell the levels shift to higher energies, which is caused by 
the increasing intrashell Coulomb repulsion between the $d$(TM) 
electrons.~\cite{Mn,Fe,Co} A comparison of Figs~\ref{fig:levels} (c) and 
(d) obtained with $U$(Ni)$=0$ and 3~eV, respectively, visualizes the 
changes induced by the $U$(TM) term. 

The possible stable charge states of TM ions are given by transition 
levels $\varepsilon$  presented in Fig.~\ref{fig:levels} (b). 
In the absence of additional dopants, a TM ion occurs in the neutral 
$q=0$ charge state, denoted as TM$^{2+}$, as long as its occupied $d$ 
levels are in the gap. In the presence of donors (acceptors), the charge 
state can change to $q=-1$ ($q=+1$), $i.e.$, to TM$^{1+}$ (TM$^{3+}$), or 
even higher ionized states.
Pronounced differences in the consecutive $\varepsilon(q/q')$  energies 
follow from the strong charge state dependencies shown in 
Figs~\ref{fig:levels} (c) and (d). 
None of defects can act as an acceptor since their $\varepsilon(0/-)$ 
levels lie above the CBM. 
In all cases except Ti, two or more charge states can be assumed. 
The $d$(Ti) levels are above the CBM not only for 
$q=0$, but also for the +1 and +2 charge state, as reflected by the 
transition level $\varepsilon(2+/+)$ being above the CBM. Therefore, a 
spontaneous autoionization of two electrons to the CBM takes place, and 
Ti occurs only in $q=+2$ charge state, in agreement with 
experiment.~\cite{Yang-Ming_2004,  Bergum, Shao_JAP2015} 
Consequently, Ti$^{4+}$ has no $d$ electrons, its spin vanishes and so 
does the $s,p-d$ coupling, and we omit Ti in the following. 
Also the V impurity has a $\varepsilon(+/0)$ level above the CBM and does 
not assume the $q=0$ charge state. 
Electron paramagnetic resonance studies of ZnO:V indicate that the stable 
charge state of V is $q=+1$,~\cite{Filipovich, Hausmann1970} in agreement 
with Fig.~\ref{fig:levels} (b). 
Other defects can occur in $q=0$, +1 and even +2 charge state depending 
on the Fermi level. 
Our results are similar to those previously reported.~\cite{Raebiger2009, 
Gluba} 

\subsection{$s,p-d$ coupling}
\begin{table}[t!]
\centering
\caption{\label{tab:spd}\small
Calculated exchange constants $N_0\alpha$, $N_0\beta_A$ and $N_0\beta_E$ 
and their average $N_0\beta$ (in eV) of 
TM$^{2+}$ and TM$^{3+}$ in ZnO.  
}
\small
\begin{tabular}{| l | r r r r r r | r r r r r r r |}
\hline 
& && $q=0$ &&&& &&& $q=1$ &&&\\
TM             & Cr  & Mn  & Fe  & Co  & Ni  & Cu 
& V & Cr  & Mn  & Fe  & Co  & Ni  & Cu\\ \hline
$N_0\alpha$    & 0.58 & 0.48 & 0.56 & 0.40 & 0.41 & 0.43
& 0.41 & 0.42 & 0.39 & 0.38 & 0.43 & 0.47 & 0.54\\
$N_0\beta_A$   & 0.14 & 0.60 & 0.80 & 4.16 & 2.32 & -5.38
& 0.18 & 0.76 & 0.48 & -0.62 & -0.91 & -1.73 & 2.67\\
$N_0\beta_E$   & 0.47 & 0.56 & 2.02 & 2.64 & 4.16 & 4.29
& 0.11 & 0.14 & 0.77 & -0.42 & -0.59 & -1.08 & -1.34\\
$N_0\beta$     & 0.36 & 0.58 & 1.61 & 3.15 & 3.55 & 1.07
& 0.13 & 0.35 & 0.67 & -0.49 & -0.69 & -1.30 & 0.00\\ 
\hline
\end{tabular}
\end{table}
\begin{figure}[t!]
\centering
\includegraphics[width=\textwidth]{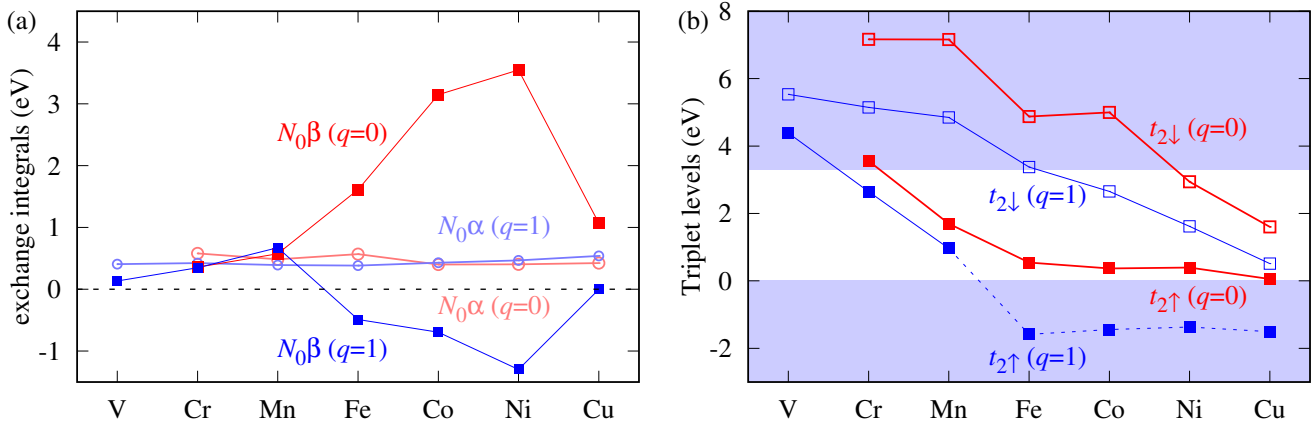}
\caption{\label{fig:spd} \small
(a) The exchange constants $N_0\alpha$ and $N_0\beta$, 
(b) the triplet levels of TM ions in ZnO for both $q=0$ and $q=+1$ charge 
states. 
The weighted average values of $N_0\beta$ and $t_{2\sigma}$ are shown. 
The resonances in the valence band are shown only schematically.
}
\end{figure}

The coupling of TM ions with the band carriers induces spin splitting of 
both the CBM, $\Delta \varepsilon_c = \varepsilon_{c\downarrow} - 
\varepsilon_{c\uparrow}$, 
and the VBM, 
$\Delta \varepsilon_{v\gamma} = \varepsilon_{v\gamma\downarrow} - 
\varepsilon_{v\gamma\uparrow}$. 
Here, $\gamma$ denotes the $A_1$ or the $E_2$ partner of the VBM. 
The exchange constants, $N_0\alpha$ for the $s-d$ coupling and 
$N_0\beta_{\gamma}$ for the $p-d$ coupling, 
are obtained directly from those splittings for supercells containing a 
single TM ion.~\cite{Sanvito} Since the splittings of 
the light hole $A_1$ and the heavy hole $E_2$ bands are 
different, we have 
\begin{eqnarray}
&&N_0\alpha=\Delta \varepsilon_c/(x \langle S \rangle), \\
\label{eq:alpha}
&&N_0\beta_A=\Delta \varepsilon_{vA}/(x \langle S \rangle),\quad
N_0\beta_E=\Delta \varepsilon_{vE}/(x \langle S \rangle),
\label{eq:beta}
\end{eqnarray}
where $\langle S \rangle= (N_{\uparrow} - N_{\downarrow})/2$, the total 
spin of the supercell, is the difference in the number of spin-up and 
spin-down electrons, $N_0$ is the density of the cation sites in ZnO, and 
$x$ is the composition of Zn$_{­1-x}$TM$_x$O ($x=0.028$ for our 
supercells). Our definitions imply that both $N_0\alpha$ and $N_0\beta$ 
are positive 
for the ferromagnetic (FM) and negative for the antiferromagnetic (AFM) 
coupling of conduction or valence $electrons$ with the TM ion. 
Note that the FM (AFM) coupling for valence electrons implies the FM 
(AFM) coupling for holes as well.

Figure~\ref{fig:spd}(a) and Table~\ref{tab:spd} shows the central result 
of this paper, namely the calculated exchange constants $N_0\alpha$, $ 
N_0\beta_A$, $N_0\beta_E$ 
and their average $N_0\beta = (1/3) (N_0\beta_A + 2 N_0\beta_E )$
for the 3$d$ TM in ZnO in both $q=0$ and $q=+1$ charge states. 
One can observe the following features that characterize the results:

\noindent 
(i) The constant $N_0\alpha$ is about 0.5~eV for all TM dopants, and 
practically does not depend on their charge state. The constant is 
positive, so the conduction electrons are ferromagnetically coupled with 
the TM impurities.~\cite{Liu_ex} 

\noindent
(ii) In contrast to $N_0\alpha$, the constant $N_0\beta$ is strongly 
dependent on the chemical identity of the dopant, as it increases 
by an order of magnitude from 0.36~eV for Cr$^{2+}$ to 3.55~eV for 
Ni$^{2+}$. 

\noindent
(iii) $N_0\beta$ can drastically depend on the impurity charge state: 
while the coupling is FM for neutral centers, it changes the 
sign to AFM for positively charged Fe, Co, and Ni ions. 

\noindent
(iv) In the case of Cu, the $p-d$ coupling with the $A_1$ and $E_2$ 
subbands is of the opposite character, resulting in opposite sign of the 
corresponding $N_0\beta_\gamma$, see Tab.~\ref{tab:spd}.
 
\section{Discussion}
\subsection{Spin splittings and the exchange-correlation potential in 
DFT}
Before a detailed discussion we present a few general remarks on the 
$s,p-d$ coupling. When the electron gas is spin polarized, the exchange-
correlation 
potential $V_{xc\sigma}$ depends on the direction of the electron spin 
$\sigma$. 
The spin splitting of the CBM is given by
\begin{eqnarray}
\label{eq:cbm_1}
\Delta \varepsilon_c = 
\varepsilon_{c\downarrow} - \varepsilon_{c\uparrow} = 
< CBM \downarrow | H^{KS}_\downarrow | CBM \downarrow > -
< CBM\uparrow | H^{KS}_\uparrow | CBM \uparrow >,
\end{eqnarray}
where $H^{KS}_{\sigma}$ is the Kohn-Sham hamiltonian with the appropriate 
$V_{xc\sigma}$. An analogous expression holds for $\Delta \varepsilon_v$. 
To simplify the discussion, we use two approximations. First, we assume 
that the orbital parts of the CBM wave functions for both spin directions 
are equal. (Validity of this approximation is discussed below.) Second, 
for illustrative purposes, we refer to the simple $X\alpha$ 
approximation, according to which the spin-dependent exchange potential 
$V_{xc\sigma}$ is given by 
$ V_{xc\sigma}(r) = A [n_{\sigma}(r)]^{1/3} $, 
where $n_{\sigma}(r)$ is the density of electrons with the spin $\sigma$ 
and 
$A=-2 e^2(3/4\pi)^{1/3}$.~\cite{Slater} 
With those assumptions, Eq.~\ref{eq:cbm_1} simplifies to 
\begin{equation}
\Delta \varepsilon_c = < CBM | \Delta V_{xc} | CBM >\qquad
\mathrm{with  }\qquad
\Delta V_{xc} = V_{xc\downarrow } - V_{xc\uparrow } =
 A(n_{\downarrow} ^{1/3} -  n_{\uparrow} ^{1/3}) 
\label{eq:cbm}
\end{equation}
and the potential difference $\Delta V_{xc}$ determines the spin 
splitting of the CBM.
 
\begin{figure}[t!]
\centering
\includegraphics[width=\textwidth]{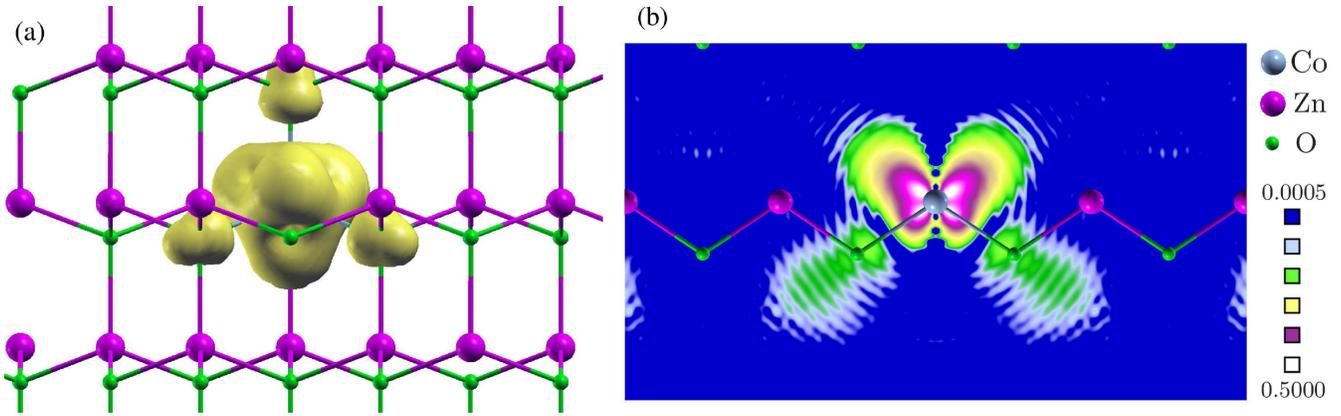}
\caption{\label{fig:spin}\small
(a) Spin density, $n_\uparrow-n_\downarrow$, of ZnO with Co$^{2+}$ (the 
isosurface value is 0.002 
electron/Bohr$^3$) and (b) two-dimensional plot of $n^{1/3}_\uparrow-
n^{1/3}_\downarrow$ in the plane containing Co-O bonds.
}
\end{figure}
According to Eq.~\ref{eq:cbm}, spin splitting $\Delta \varepsilon_c$ 
is given by the product of $\Delta V_{xc}$ and the wave function squared. 
An insight into the mechanisms of the $s,p-d$ coupling is obtained by 
inspecting those quantities. 
The first factor, $\Delta V_{xc}$, is a functional of the densities 
$n_{\sigma}$, and stems from the non-vanishing spin density $\Delta n= 
n_{\uparrow} - n_{\downarrow}$. It is mainly localized on the $d$-shell 
of the TM ion. The final charge and spin densities can be considered as a 
result achieved in two steps. First, in the absence of coupling between 
the ZnO host and, e.g., the Mn dopant, the ZnO:Mn system consists in the 
ZnO host with one vacancy and the Mn atom. The total spin of this system 
is 5/2. After switching on the coupling, $i.e.$, the $p-d$ hybridization, 
the Mn spin "spills" onto the ZnO host, mainly onto the first O neighbors 
of Mn. Thus, the Mn spin is somewhat reduced and the O neighbors become 
spin polarized, but the total spin is conserved and equal 5/2. 
The final spin polarization acts as a source of an additional attractive 
potential for the spin-up electrons, which determines the FM character of 
the coupling and causes the spin splitting of the band states. 

$\Delta n$ is shown in Fig.~\ref{fig:spin} (a) for Co in ZnO. It is 
dominated by the Co orbitals, and also contains a contribution from the 
spin polarized O nearest neighbors. Figure~\ref{fig:spin} (b) shows the 
difference ($n_{\uparrow} ^{1/3} - n_{\downarrow} ^{1/3}$), to which 
$\Delta V_{xc}$ is approximately proportional. The shape of the 
isosurface has a tetrahedral symmetry to a good approximation, and 
reflects that of $\Delta n$. Co$^{2+}$ has 7 $d$-electrons. From Fig. 
\ref{fig:levels} (a) it follows that the two $e_2$ orbitals are occupied 
with 4 electrons, two with spin-up and two with spin-down, and their 
total spin is zero. The spin density shown in Fig.~\ref{fig:spin} is thus 
dominated by the three $t_{2\uparrow}$ orbitals. 
Because $V_{xc\sigma}$ is mainly given by the (1/3) power of 
$n_{\sigma}$, $\Delta V_{xc}$ is smoother and somewhat more delocalized 
than the spin density, which enhances the role of the O neighbors.
\begin{figure}[t!]
\centering
\includegraphics[width=\textwidth]{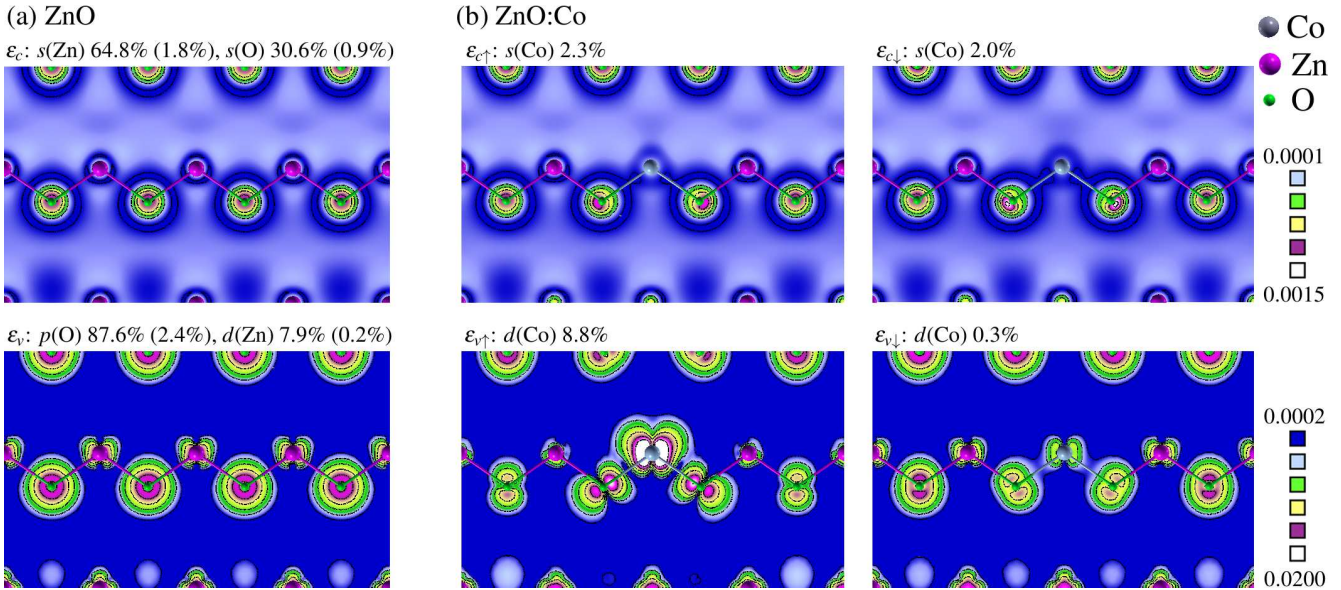}
\caption{\label{fig:wf}\small
(a) The wave functions squared of the CBM and 
the VBM (sum over $p$ states) of ZnO and their orbital compositions.
The numbers give the total contribution of orbitals, and in parentheses 
we give the contribution per one cation or one anion in the 72-atom 
supercell (e.g., $1.8\% \times 36 = 64.8\%$).
(b) The wave functions squared of the CBM and the VBM of ZnO:Co$^{2+}$.
Left (right) panels show the spin-up (spin-down) states. 
The contributions of $s$(Co) or $d$(Co) orbitals are shown in each case.}
\end{figure}

The second factor which determines the spin splitting of a given state is 
its wave function. In a zinc blende crystal, the CBM of the $\Gamma_1$ 
symmetry contains the $s$ orbitals only, while the $\Gamma_{15}$ VBM 
states are composed of both $p$ and $d$ orbitals. In the wurtzite ZnO 
these selection rules are relaxed by the hexagonal part of the crystal 
field, but this small effect can be neglected in the discussion. 
Those features are seen in Fig.~\ref{fig:wf} (a), which presents the wave 
functions of pure ZnO as a reference. In the case of ZnO:Co,  
Fig.~\ref{fig:wf} (b), hybridization between the TM and the host ZnO 
states results in the contribution from $s$(TM) to the CBM, and from  
$d$(TM) to the VBM, as it is discussed in detail below. 
Note that the effect of $p-d$ hybridization on the $t_{2\sigma}$ and 
$e_{2\sigma}$ gap states is even more pronounced. 
It is reflected in their strong delocalization 
shown in Fig.~\ref{fig:TMwf} of Supplementary Information, especially when 
compared with the compact spin polarization $\Delta n$ shown in 
Fig.~\ref{fig:spin}.

Our results provide a good illustration of the fact that the Heisenberg 
form of the coupling, $H_{ex}=-J\ {\bf S\cdot s}$, has an "effective" 
character. Indeed, the Heisenberg hamiltonian suggests that the exchange 
interaction acts on the spin component of the wave function. Actually, 
this is not the case, because $V_{xc\sigma}$ is spin-dependent but it 
acts on the orbital parts of the wave function. It leads to the 
differences between the spin-up and spin-down partners of CBM and VBM, 
and the main difference relies in the different contribution of the 
$d$(TM) to the wave functions shown in Fig.~\ref{fig:wf_S} of Supplementary 
Information.  

\subsection{$s-d$ exchange coupling}
Theory of the exchange coupling between the $s$-like conduction electrons 
and the localized $f$-shell in rare earths (REs) was developed by 
Liu.~\cite{Liu_ex} 
He considered a restrained set of states, namely the $f$-shell of the RE 
atom and the CBM, and applied the Hartree-Fock approach to obtain the 
exchange term. 
This direct (or potential) exchange stems from the overlap of 
the corresponding wave functions, and is FM. 
Because of the strong localization of the $f$ orbitals, the overlap 
integrals extend over the volume of the magnetic atom only, and therefore 
the $s-f$ coupling is driven by intra-atomic effects. 
The Liu's picture was then applied to the $s-d$ coupling in DMSs 
in spite of differences between the two 
systems, which we discuss in the following. 

The tight binding picture allows for an intuitive real space 
interpretation of the results at the atomic level. 
The CBM wave functions are represented as sums over the appropriate $s_R$ 
orbitals of atoms at sites $R$ in the supercell: 
\begin{equation}
\psi_{\sigma}(r) = \sum_{R} a_{R\sigma}  s_R(r-R),   
\label{eq:psi}
\end{equation}
and are normalized to 1 in the supercell volume. 
The decomposition coefficients $a_{R\sigma}$ are obtained by projection 
of the calculated $\psi_{\sigma}$ onto the atomic orbitals. 
Assuming again that the orbital part of the wave function is the same for 
both spin directions we have an approximate expression 
\begin{eqnarray}
\Delta \varepsilon_c =  a_{R=0}^2 <s_{TM} | \Delta V_{xc} | s_{TM} > 
+\sum_{R=NN} a_{R}^2 < s_O | \Delta V_{xc} | s_O > 
= \Delta \varepsilon_c(TM)  + \Delta \varepsilon_c(O_{NN}), 
\label{eq:split1}
\end{eqnarray}
\noindent
\noindent
where the sum is limited to the first oxygen neighbors of the TM ion, 
in agreement with the strong localization of both $\Delta n$ and 
$\Delta V_{xc}$, see  Fig.~\ref{fig:spin}. 
In this expression the inter-atomic terms are neglected, which is 
justified by the smallness of the involved overlap integrals, and the 
spin 
splitting of the CBM is the sum of atomic-like contributions. 
The first one, $\Delta \varepsilon_c(TM)$, corresponds to the Liu picture 
of the $s-d$ coupling, while the second one, $\Delta 
\varepsilon_c(O_{NN})$, originates in the spin polarization of the O ions 
induced by the $p-d$ hybridization with $d$(TM). 

To estimate the first term of Eq.~\ref{eq:split1} one needs the splitting 
energy $\Delta_{spin}^{atom}$ of the $s_{TM}$ orbitals of isolated atoms. 
Unfortunately, as it is explained in the Supplementary Information, 
$\Delta_{spin}^{atom}$ can be evaluated only for Mn, because there are 
fundamental problems with obtaining a correct electronic structure of the 
remaining TM atoms within the DFT. 
In the case of Mn, the value of the first term of Eq.~\ref{eq:split1}  is 
obtained under two assumptions. First, the spin polarization is entirely 
localized on the TM ion, which corresponds to the Liu's picture. The 
calculated $\Delta_{spin}^{atom}$(Mn)$=1.0$~eV. Second, the difference 
between $a_{TM\uparrow}^2=0.025$ and $a_{TM\downarrow}^2=0.018$ is 
neglected, and the average value $a_{R=0}^2=<a_{TM\sigma}^2>=0.021$ is 
used instead. This gives $\Delta \varepsilon_c$(Mn)$ =  a_{R=0}^2 
\Delta_{spin}^{atom} = 0.021$~eV, which is about (2/3) of the actual 
splitting of 0.034~eV. This estimation is likely to represent the upper 
limit, since $\Delta V_{xc}$ in ZnO is more delocalized, and thus weaker, 
than in an isolated Mn atom.  
Moreover, this shows that the second term of Eq.~\ref{eq:split1}, $\Delta 
\varepsilon_c(O_{NN})$ contributes about 1/3 to the CBM spin splitting. 
Thus, the $p-d$ hybridization plays an indirect but non-negligible role 
in the $s-d$ exchange coupling, leading to the spin-polarization not only 
of $p$(O) but also of $s$(O) electrons and thus enhancing the values of 
$N_0\alpha$.  
A comparable situation is expected to take place for other TM ions given 
the similarity in the CBM wave functions and in spin polarization of the 
O anions.

As follows from Fig.~\ref{fig:spd} (a), $N_0\alpha$ is almost independent 
of the dopant and its charge state, especially when compared with large 
changes in $N_0\beta$. 
Actually, the deviations from the average value of 0.5~eV are about 20 
per cent.
This result can be related with the intra-atomic character of the $s-d$ 
coupling. We first note that the definition of $N_0\alpha$ through 
$\Delta \varepsilon _c=-N_0\alpha\ \bf{S\cdot s}$ implies that $N_0\alpha$ 
describes the coupling between a free carrier and $one$ of the $d$(TM) 
electrons. This in turn is given (in the Hartree-Fock picture) by the 
exchange overlap integral between the $s$(TM) and $d$(TM) orbitals of the 
dopant. Those integrals should be similar within the 3$d$ series because 
of the similarity of the involved $s$ and $d$ states. Also, both $s$ and 
$d$ are not expected to strongly depend on the charge state. Second, the 
decomposition of the CBM wave functions shows that the contributions from 
the $s$(TM) orbitals to the CBM are similar for all dopants, about 1.5-
2.5\% (in the 72-atom supercells), with no clear trend regarding the TM 
identity or charge state. These two factors combined contribute to the 
obtained weak dependence of $N_0\alpha$ on the dopant.

\subsection{$p-d$ exchange coupling}
\begin{figure}[t!]
\centering
\includegraphics[width=0.5\textwidth]{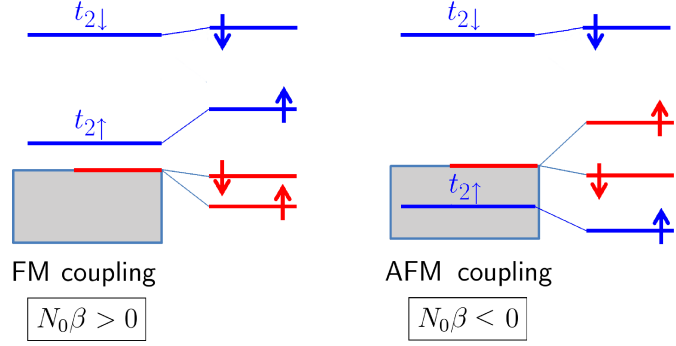}
\caption{\label{fig:pd} \small
Dependence of the $p-d$ exchange coupling on the energies of the 
$t_{2\sigma}$(TM)  levels. 
The hybridization occurs between the states with the same spin, and is 
spin-dependent. Left 
panel: both $t_{2\uparrow}$ and $t_{2\downarrow}$ are above the VBM. 
$t_{2\uparrow}$ is closer to the VBM than $t_{2\downarrow}$, and thus its 
interaction with 
the VBM is stronger. The resulting spin splitting corresponds to the FM 
$p-d$ exchange 
coupling. Right panel: $t_{2\uparrow}$ is below the VBM, and the 
$t_{2\downarrow}$ is 
above the VBM, which results in the AFM coupling. Gray area is the 
valence band. Arrows 
indicate spin direction of levels only, and occupations are not shown. 
Left panel represents e.g. Mn 
in ZnO, and the right one Mn in CdTe.   Crystal field splittings of the 
triplets are neglected for 
simplicity. }
\end{figure}
Hybridization between the VBM and the $d$(TM) states is essential for the 
$p-d$ coupling. It is typically analyzed in the second 
order of perturbation theory, in which  $\Delta \varepsilon_v$ of a cubic 
semiconductor is determined by the energies of the TM levels 
$\varepsilon(t_{2\sigma})$ relative to the VBM~\cite{Sato}
\begin{equation}
\Delta \varepsilon_v  =
\frac{1}{2}\left(
{|V_{hop,\uparrow }|^2\over \varepsilon(t_{2\uparrow}^0) - 
\varepsilon_v^0}-{|V_{hop,\downarrow}|^2\over 
\varepsilon(t_{2\downarrow}^0) - \varepsilon_v^0}
\right), 
\label{eq:vb}
\end{equation}
where superscript indexes "0" mean the unperturbed $d$ and VBM level 
energies, and $V_{hop,\sigma}$ is a spin dependent hopping integral 
between a TM ion and its neighbors. 
Average energies of the TM triplet levels are shown in 
Fig.~\ref{fig:spd} (b). 
Although the Figure presents the final self-
consistent energies $\varepsilon(t_2)$ rather than $\varepsilon(t_2^0)$, 
they can serve as a basis for discussion. Equation~\ref{eq:vb} together 
with Fig.~\ref{fig:spd} (b)  qualitatively explain the calculated 
characteristics of $N_0\beta$ in terms of the $d$-shell energy levels 
relative to the VBM. 
For the neutral TM dopants, 
the first term of Eq.~\ref{eq:vb} gives a positive ($i.e.$, FM), while 
the second one gives a negative (AFM) contribution to $N_0\beta$. 
The first term is dominant, and $N_0\beta$ 
is positive since $t_{2\uparrow}$ is closer to the VBM than 
$t_{2\downarrow}$. With the increasing atomic number, $N_0\beta$ 
increases due to the decreasing energy denominators, see 
Fig.~\ref{fig:spd} (a). 
In turn, $t_{2\uparrow}$ level for positively charged Fe, Co and Ni is 
$below$ the VBM and thus both terms of Eq.~\ref{eq:vb} lead to AFM 
exchange coupling. 
The underlying mechanism is schematically shown in Fig.~\ref{fig:pd}. 

\begin{figure}
\centering
\includegraphics[width=\textwidth]{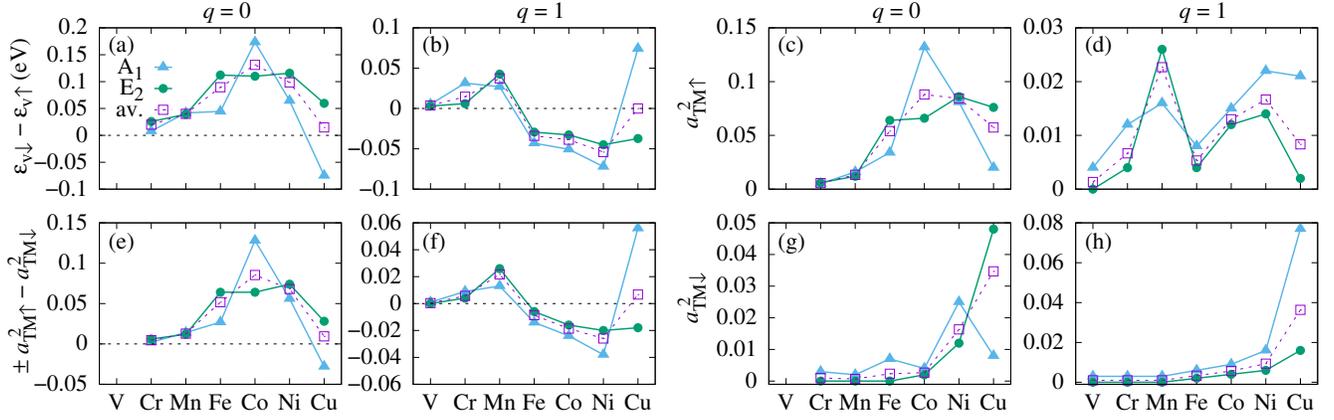}
\caption{\label{fig:correl}\small
(a, b) The spin splitting of the singlet $A_1$ and the doublet $E_2$ hole 
states together with its weighted average, (c, d) and (g, h) the 
appropriate decomposition coefficients $a_{TM\uparrow}^2$ and 
$a_{TM\downarrow}^2$ of the VBM functions and (e, f) their combination 
$\pm a_{TM\uparrow}^2-a_{TM\downarrow}^2$, see details in the text. 
Charge states $q$ are indicated on the top.
}
\end{figure}

In the case of ZnO, the situation is somewhat more complex, since the VBM 
is a quasi-triplet formed by the $A_1$ and the $E_2$ hole subbands. The 
corresponding TM-induced spin splitting energies are given in 
Fig.~\ref{fig:correl} (a) and (b). 
According to our results, 
$\Delta\varepsilon_A$ and $\Delta\varepsilon_E$ can 
substantially differ, and the difference critically depends on both the 
dopant and its charge state. 
In particular, the spin density of the Mn$^{2+}$ with the fully occupied 
spin-up $d$-shell is a fully symmetric $\Gamma_1$ object, it acts 
on $A_1$ and $E_2$ in a very similar fashion, and consequently 
$\Delta\varepsilon_A$ and $\Delta\varepsilon_E$ are almost equal. This is 
not the case of other TM ions. They are characterized by "non-spherical",
$i.e.$, non-$\Gamma_1$, 
spin densities, which act differently on the $A_1$ and $E_2$ partners, 
and induce different $\Delta\varepsilon_A$ and $\Delta\varepsilon_E$. For 
Cu, even the signs of the splittings are opposite. 

The impact of the $p-d$ hybridization on the $p-d$ coupling is well 
illustrated by Fig.~\ref{fig:correl}. The Figure shows the decomposition 
coefficients $a_{TM\sigma}^2$ of the VBM wave functions in the tight 
binding picture analogous to Eq.~\ref{eq:psi}. 
By comparing Figs~\ref{fig:correl} (c)-(d) with (g)-(h) one observes 
that the hybridization is strongly spin-dependent, since
the contribution of the spin-up and spin-down TM states to the VBM can 
differ by as much as one order of magnitude. This stems from the 
different energies of the TM gap states relative to the VBM, $i.e.$, the 
different energy denominators in Eq.~\ref{eq:vb}, which also control the 
mixing of wave functions. 
In most cases, the spin-down states are more distant from the VBM than 
the spin-up ones, and the contribution of $d$(TM) to the VBM is 
appreciably larger for the spin-up than for the spin-down channel. 
However, since the spin splitting is the energy difference, we have to 
consider 
the appropriate combinations of the $a_{TM\uparrow}^2$ and 
$a_{TM\downarrow}^2$ coefficients rather than their values separately. 
Depending on the actual level ordering, the combination is 
$\pm a_{TM\uparrow}^2-a_{TM\downarrow}^2$, consistently with the 
Eq.~\ref{eq:vb} , 
where the $+ (-)$ sign holds when the spin-up TM level is above (below) 
the VBM. 
The results are shown in Fig.~\ref{fig:correl} (e)-(f). 
The very high level of correlation between the splittings 
$\Delta\varepsilon_A$ and $\Delta\varepsilon_E$ and the contribution of 
the $d$(TM) orbitals to the VBM is clear. 
We also note that because of the large differences between 
$a_{TM\uparrow}^2$ and $a_{TM\downarrow}^2$, the approximate 
Eq.~\ref{eq:split1} cannot be applied to the VBM, and the appealing 
separation into the Liu intra-atomic contribution and the hybridization 
contribution does not hold. 
However, the intra-atomic contribution of $d$(TM) to 
$\Delta\varepsilon_v$ does not vanish. 
Since the direct exchange leads to the FM coupling, it enhances the 
hybridization-induced values of the $N_0\beta$s for all TM$^{2+}$ 
dopants, and reduces the negative values of $N_0\beta$s for Fe$^{3+}$, 
Co$^{3+}$ and Ni$^{3+}$.

In a complementary approach, one can follow the impact of the $p-d$ 
hybridization in the real space picture. The relevant wave functions for 
ZnO:Co are given in Fig.~\ref{fig:wf}. In agreement with the attractive 
character of $\Delta V_{xc}$ and with Fig.~\ref{fig:correl},  the spin-up 
VBM wave function contains a larger contribution of $d$(Co) than its 
spin-down partner. Indeed, from Fig.~\ref {fig:wf} (b) it follows that 
the approximation of Eq.~\ref{eq:cbm} is not justified for the VBM, 
because the contribution of Co  to the spin-up (8.8\%) and spin-down 
states (0.3\%) differ over 20 times. The contributions of the $d$(TM) 
orbitals to the VBM for other TM dopants is shown in Fig.~\ref{fig:wf_S} of 
Supplementary Information. They depend on the dopant and its charge 
state. 

Analyzing the consecutive TM ions we find that, as it follows from 
Fig.~\ref{fig:correl}, in the case of the light V the contribution of the 
$d$(TM) shell to the VBM practically vanishes. Accordingly, the VBM spin 
splittings and the corresponding $N_0\beta$s are small. Next, the average 
values of $N_0\beta$ increase to 0.4 -- 0.6~eV for Cr and Mn with the 
decreasing energies of $t_2$. In the sequence Fe -- Ni, $t_{2\uparrow}$ 
is very close the valence band, which strongly enhances both the spin 
splitting of the VBM and $N_0\beta$. Finally, Cu represents an 
interesting case, since the splittings $\Delta\varepsilon_A$ and 
$\Delta\varepsilon_E$ have opposite signs. This effect takes place 
because some of the energies of the $d$(Cu) states are below the VBM, 
thus changing the sign of the splitting in agreement with 
Fig.~\ref{fig:correl}. 
          
In the case of charged dopants with $q=+1$, the TM-induced levels are 
lower in energy than for $q=0$, see Fig.~\ref{fig:spd} (b). This leads to 
higher values of $N_0\beta$ for Cr and Mn. More importantly, in the case 
of Fe, Co, and Ni, the $t_{2\uparrow}$ levels are $\it below$ the VBM, 
which changes the sign of the first term in Eq.~\ref{eq:vb}, and drives 
the change of character of the $p-d$ coupling to AFM, as displayed by the 
negative $N_0\beta$ in Fig.~\ref{fig:spd} (a). In particular, the AFM 
coupling can be expected for Fe, which typically assumes  the Fe$^{3+}$ 
charge state.~\cite{Fe}

\subsection{Comparison with previous calculations and with experiment}
The role of the $p-d$ hybridization in the $p-d$ exchange coupling was 
recognized early.  Typically, it was taken into account by using the 
Anderson hamiltonian.~\cite{Anderson} 
The calculations employing the Schreiffer-Wolf\cite{Schrieffer} 
transformation were
performed and found a negative $N_0\beta$ for Mn in II-VI semiconductors 
like CdTe.~\cite{Larson, Zunger1987} Indeed, it was properly recognized 
that the $d$(Mn) states in CdTe are placed about 2 eV below the VBM. As 
it is shown in Fig.~\ref{fig:pd}, in this situation the $p-d$ exchange 
coupling of holes with the Mn$^{2+}$ spins must be AFM independent of the 
details of calculations. This approach was used for other TM ions leading 
to AFM coupling for Mn$^{2+}$, Fe$^{2+}$, Co$^{2+}$~\cite{Bhattacharjee, 
Blinowski, Kacman} and FM coupling for Sc$^{2+}$ and 
Ti$^{2+}$.~\cite{Blinowski, Kacman}
The configuration interaction and cluster-model calculations were also 
used to evaluate $N_0\beta$ for several II-VI hosts and TM dopants, and 
to interpret the experimental data; the obtained $N_0\beta$'s were 
negative indicating the AFM coupling with holes.~\cite{Mizokawa1993, 
Mizokawa1997, Mizokawa2002, Okabayashi} 
In the papers above, simple expressions for $N_0\beta$ are given, in 
which the critical factor is the energy of the majority spin $d$(TM) 
level relative to the VBM. 
In the case of ZnO, the TM level energies were assumed to be similar to 
those in CdTe, which leads to AFM coupling with $N_0\beta$ of about $-
3$~eV for Mn$^{2+}$, Fe$^{2+}$ and Co$^{2+}$.~\cite{Blinowski2001, 
Beaulac,Okabayashi}

However, the assumption that $t_{2\uparrow}$(Mn) is below the VBM in ZnO 
was invalidated by experiment. The measurements~\cite{Johnson} proved 
that $t_{2\uparrow}$(Mn) is in the band gap, which was subsequently 
confirmed by both experiment and theory.~\cite{Gilliland, Raebiger2009, 
Mn} Consequently, in this case the $p-d$ coupling is FM. This should be 
the case of Cr as well, since $t_{2\uparrow}$(Cr) is higher than that of 
Mn. 
Similarly, the energy of the gap levels of Co$^{2+}$ are well 
established,~\cite{Raebiger2009,Co} and $N_0\beta>0$ is expected. This 
may imply that the interpretation of the X-ray data, leading to the 
negative $N_0\beta$ for Mn,\cite{Mizokawa2002, Okabayashi} and possibly 
for Fe, Co, and Ni, was not correct.  
Finally, as observed in Ref.~\cite{Johnson}, "location of the $d$ levels 
below the VBM is an essential assumption behind the proposal of mid-gap 
Zhang-Rice-like states in ZnO:Mn.~\cite{Dietl}" According to the results 
presented above this assumption is not correct. 

The same issue, $i.e.$, correctness of energies of the $t_2$(TM) levels 
relative to the CBM and VBM, is present in the case of calculations based 
on the local density approximation. This approximation results in a 
severe underestimation of the ZnO band gap, and therefore wrong energies 
of the TM levels and the non-correct sign of the $p-d$ coupling (e.g., 
$N_0\beta=-1.81$~eV for Mn in ZnO~\cite{Chanier}). On the other hand, 
when the correct $E_{gap}$ of ZnO is used,~\cite{Raebiger2009} the 
energies of the TM gap states are close to the present results. 

We now turn to the experimental results. In early magneto--optical 
measurements, a strong exchange interaction between band carriers and 
localized $d$(TM) electrons was reported for Mn, Fe, Co, Ni and 
Cu.~\cite{Ando:APL,Ando:JAP2001,Ando:JPCM2004,Kataoka} By contrast, the 
$s,p-d$ coupling was not observed so far for Ti, V and Cr ions in 
ZnO.~\cite{Ando:JAP2001} This latter result is in a reasonable agreement 
with our findings. Indeed, the stable charge state of Ti is the 
nonmagnetic Ti$^{4+}$ ($q=2$) as in experiment,~\cite{Yang-Ming_2004,  
Bergum, Shao_JAP2015} and thus both exchange constants, $N_0\alpha$ and 
$N_0\beta$, vanish by definition. Also, V (stable in the V$^{3+}$ charge 
state according to both experiment~\cite{Filipovich, Hausmann1970} and 
our calculations), as well as Cr, are characterized by small $N_0\beta$  
exchange constants. 

Subsequent and more detailed experiments were performed for Mn, Fe, Co 
and Ni. In the case of Mn$^{2+}$, $N_0|\beta-\alpha|=0.2\pm 0.1$~eV was 
determined by magneto-optical measurements, which gives $N_0\beta=0.5\pm 
0.2$~eV or $N_0\beta=0.1\pm 0.2$~eV  with the assumption that 
$N_0\alpha=0.3\pm 0.2$~eV.~\cite{Pacuski:PRB2011} Similar values, 
$N_0|\alpha-\beta|=0.1$~eV ~\cite{Przezdziecka_ssc} and $N_0|\alpha-
\beta|=0.16$ ~eV~\cite{Przezdziecka_pssc}, were also determined. These 
results are in a good agreement with the calculated $N_0(\beta-
\alpha)=0.1$~eV. 

In the case of Co, magnetooptical measurements showed that $N_0|\beta-
\alpha|=0.8$~eV.~\cite{Pacuski:PRB2006} The sign of $N_0\beta$ could not 
be determined experimentally due to the ambiguity in the valence bands 
ordering.
By assuming $N_0\alpha=0.25$~eV, it was concluded that the $p-d$ coupling 
can be either FM with $N_0\beta=1.0$~eV, or AFM with $N_0\beta=-0.55$~eV. 
Our calculated $N_0\beta$ depends on the charge state. For the neutral 
Co$^{2+}$ we find the FM coupling ($N_0\beta_\gamma = 4.2 $ and 2.6~eV 
for the $A_1$ and $E_2$ subbands, respectively), while for the positively 
charged Co$^{3+}$ the coupling is AFM ($N_0\beta_\gamma = -0.9$ and $-
0.6$~eV for the $A_1$ and $E_2$ subbands, respectively). Thus, we obtain 
that $N_0|\beta-\alpha|=2.6$~eV for Co$^{2+}$, and about 1.2~eV for 
Co$^{3+}$. The latter value is reasonably close to magnetooptical 
data.~\cite{Pacuski:PRB2006} 

$N_0\beta$s determined for Fe and Ni are large and 
negative, namely -2.7~eV~\cite{Okabayashi}  and $-4.5\pm 
0.6$~eV,~\cite{Schwartz} respectively. These values are somewhat higher 
than our results for $q=0$ ($N_0\beta_E=2.0$ and $N_0\beta_A=0.8$ ~eV for 
Fe, and $N_0\beta_E=4.2$ and $N_0\beta_A=2.3$ ~eV for Ni), but 
importantly they are of opposite sign. However, we predict negative but 
smaller values for those dopants in the $q=+1$ charge states. 
Here, it should be mentioned that large and negative $N_0\beta$s were 
proposed also for Mn (--3.0~eV,~\cite{Okabayashi} --2.7~\cite{Schwartz})  
and for Co (--3.4~eV,~\cite{Okabayashi} --2.3~\cite{Schwartz}).
However, as it is pointed out above, interpretation of these measurements 
is based on particular assumptions regarding the energies of the TM 
levels, which were subsequently questioned. We conclude that a reliable 
comparison with experiment for Fe, Co, and Ni requires the charge state 
of the TM ion to be established. 

\section{Summary and Conclusions}
Theoretical analysis of the $s,p-d$ exchange coupling between free 
carriers and the $3d$ transition metal dopants in ZnO  was conducted 
employing the GGA$+U$ method. The present study reveals both the detailed 
characteristics for each ion, and general trends. A particular care was 
devoted to reproduce the correct band gap of ZnO and energies of the gap 
levels of the dopants. 
The calculated $s-d$ coupling constant $N_0\alpha$ is about 0.5~eV for 
all the TM ions, $i.e.$, it does not depend on the dopant and its charge 
state. By contrast, the $p-d$ exchange coupling reveals unexpectedly 
complex features. First, $N_0\beta$s strongly depend on the chemical 
identity of the dopant, increasing about 10 times from V to Cu. Second, 
$N_0\beta$ is different for the two VBM subbands, the light hole $A_1$ 
and the heavy hole $E_2$, and the corresponding values can differ by a 
factor 2, or even have opposite signs. Third, not only the magnitude but 
also the sign of $N_0\beta$ depends on the charge state of the TM ion. In 
particular, the coupling between holes and Fe, Co and Ni ions in the 
$q=0$ neutral charge state is strong and ferromagnetic, while for $q=+1$ 
the coupling changes the character to antiferromagnetic. 
Finally, the stable charge state of Ti in ZnO is Ti$^{4+}$, in which  its 
spin vanishes. 

Analysis of the wave functions reveals how the hybridization between the 
TM orbitals and the ZnO band states determines the $s,p-d$ exchange 
coupling. The magnitude of the $p-d$ coupling is determined by the 
energies of the $d$(TM) relative to the VBM. The most striking example is 
that of Cu, for which the $t_{2\uparrow}$(Cu) and VBM are almost 
degenerate, and the actual ordering of the $d$(Cu)-induced levels and the 
VBM explains different signs of $N_0\beta$s of the light and the heavy 
holes, and their large magnitudes. Thus, the $p-d$(TM) hybridization 
leads to the Anderson-like picture of the $p-d$ coupling, but its role is 
more complex. 
In particular, the $p-d$ hybridization affects not only $N_0\beta$ but 
also the $N_0\alpha$ constant. The main mechanism of the $s-d$ coupling 
is grasped by the Liu's model, and it originates in the TM intra-atomic 
exchange interaction between the $s$ and $d$ electrons. 
However, the spin polarization of the oxygen neighbors of the TM ion 
induced by the $p-d$ hybridization leads to the spin polarization of the 
$s$(O) orbitals, which contributes about 1/3 to the $N_0\alpha$ constant. 

Comparison with experiment is satisfactory for Ti, V, Cr and Mn. In the 
case of Fe and Co, the definitive conclusions are not possible, because 
$N_0\beta$ depends on the dopant charge state, which were not assessed in 
experiment. An acceptable agreement for Co is obtained assuming the 
Co$^{3+}$ and not the Co$^{2+}$ charge state.

\section{Method of calculations}
The calculations are performed within the density functional 
theory~\cite{Hohenberg,KohnSham} in 
the generalized gradient approximation (GGA) of the exchange-correlation 
potential $V_{xc}$,~\cite{PBE} supplemented by the 
$+U$ corrections.~\cite{Cococcioni} 
We use the pseudopotential method 
implemented in the {\sc Quantum ESPRESSO} 
code,~\cite{QE} 
and employ ultrasoft pseudopotentials, which include nonlinear core 
correction in the case of Co and Ni.
The valence atomic configuration is $3d^{10}4s^2$ for 
Zn, $2s^2p^4$ for O, and $3s^2p^6 4s^2p^0 3d^n$ or $4s^2p^0 3d^n$ for TM 
ions with $n$ electrons on the $d$ shell. 
For V, Ti, Cr, Mn, Fe and Co, the plane-waves kinetic energy cutoffs 
of 30~Ry for wave functions and 180~Ry for charge density are employed. 
Convergence was assessed by test calculations with cutoffs of 40~Ry. 
Following the recommendation of {\sc Quantum ESPRESSO}, for Ni and Cu the 
cutoff is increased to 45~Ry. 

Spin-orbit interaction is neglected. We justify this approximation 
by the results of experiments regarding TM dopants in ZnO. The 
interaction manifests itself in optical measurements, where the lines of 
intracenter $d-d$ transitions reveal rich structures, and are split in 
particular by the spin-orbit coupling. According to the results for 
Co,~\cite{Koidl, Schulz} the spin-orbit splittings of the initial and 
final states of the T$^4_2$(F) – A$_2$(F) emission can be estimated as 
19 and 6 cm$^{-1}$, respectively, which corresponds to 1-3~meV. Similar 
values, lower than 10~meV, were reported for other TM 
dopants.~\cite{Schulz, Malguth} In the case of Mn$^{2+}$ and Fe$^{3+}$, 
due to the absence of orbital momentum in the $d^5$ configuration, the 
second order spin–orbit interaction leads to splitting energies below 
1~meV.~\cite{Malguth} Moreover, the spin-orbit splitting of the VBM in 
ZnO is about 10~meV, $i.e.$, it is very small compared to the spin 
splittings of the order of 1~eV characterizing heavier atoms and 
semiconductors such as InSb, CdTe or PbTe. The smallness of the spin-
orbit coupling in ZnO:TM justifies its neglect both in our~\cite{Mn, Fe, 
Co} and in the previous ab initio calculations.~\cite{Raebiger2009,Gluba} 

The electronic structure of the 
wurtzite ZnO is examined with an $8\times 8\times 8$ $k$-point grid. 
Analysis of a single TM impurity in ZnO is performed using $3\times 
3\times 2$ supercells with 72 atoms, while $k$-space summations are 
performed with a $3\times 3\times 3$ $k$-point grid. 
For too small supercells, the spurious defect-defect coupling can distort 
final results. Convergence of the results with respect to the supercell 
size was checked for Mn and Co with $6\times 6\times 4$ supercells with 
576 atoms. 
We obtained that the $N_0\alpha$s are the same 
for both supercells, while $N_0\beta$s are lower by 10 per cent for Mn 
and 20 per cent for Co in the case of the larger supercell. 
Ionic positions are optimized until the forces acting on ions became 
smaller than 0.02~eV/\AA.

The parameters $U(\textrm{Zn})=12.5$~eV for $3d(\textrm{Zn})$ and 
$U(\textrm{O})=6.25$~eV for $2p(\textrm{O})$ electrons are fitted to 
reproduce the experimental ZnO band gap $E_{gap}$ of 3.3~eV,~\cite{Dong, 
Izaki:APL1996, Srikant:JAP1998} the width of the upper valence band of  
6~eV and the energy of the $d$(Zn)-derived band.~\cite{Lim} 
Our $U$ values are similar to those reported in other works.~\cite{Ma, 
Calzolari, marco} The lattice parameters $a= 3.23$~(3.25)~\AA, $c = 
5.19$~(5.20)~\AA\ and $u = 0.38$~(0.38) are underestimated by less than 
1~\% in comparison with experimental values~\cite{Karzel} given in 
parentheses.

The used $U$ corrections for $3d(\textrm{TM})$ electrons 
are: $U$(Ti)=2.0~eV, $U$(V)=2.0~eV, $U$(Cr)=2.0~eV, $U$(Mn)=1.5~eV, 
$U$(Fe)=4.0~eV, $U$(Co)=3.0~eV, $U$(Ni)=3.0~eV, and $U$(Cu)=2.0~eV. 
For Mn, Fe, Co and Cu they were optimized by a careful fitting to the 
experimental energies of both intra-center and ionization optical 
transitions.~\cite{Mn,Fe,Co,Cu} For other TM dopants, the 
$U$ corrections are taken to be 2-3~eV, as suggested in the 
literature.~\cite{Gluba,Raebiger2009} 
We checked that in most cases a variation of the $U$(TM) value by 1~eV 
alters the impurity levels by about 0.1~eV, and the 
$N_0\alpha$ or $N_0\beta$ values by less than 0.05~eV. 
We also mention that in spite of the energetic proximity and strong 
hybridization between the VBM and the TM-induced levels, a non-ambiguous 
identification of hole states was always possible based on the analysis 
of wave functions. However, since the $p-d$ coupling depends on the 
inverse energy distance between the TM-induced levels and the VBM, the 
results are less accurate for Co, Ni and Cu than for Cr and Mn. 

Various charge states of the TM dopants are considered. In general, in 
the absence of additional dopants, a TM ion occurs in the neutral $q=0$ 
charge state, denoted as TM$^{2+}$, and other charge states $q$ can also 
be assumed when defects are present. 
Generally, the stable charge state of a defect in a semiconductor depends 
on the Fermi level. Transition level $\varepsilon(q/q')$ of a defect is 
defined as the Fermi energy at which the stable charge state changes from 
$q$ to $q'$, or in other words, as the Fermi energy at which formation 
energies of $q$ and $q'$ are equal:
\begin{equation}
\label{eq:trans_level}
\varepsilon(q/q')=\frac{E(q')-E(q)}{q-q'}-\varepsilon_v,
\end{equation}
where $E(q)$ is the total energy of the doped supercell and 
$\varepsilon_v$ is the VBM energy of pure ZnO. 
The finite size effects are taken into account by including the image 
charge corrections and
potential alignment for charged 
defects.~\cite{Lany_PRB78,Lany_ModelSim17}
The energies $\varepsilon(q/q')$ 
in the gap determine the possible charge states of TM ions. 

Finally, the spin splitting energies of the VBM and CBM are taken 
directly from the Kohn-Sham levels. Alternatively, they can be obtained 
from appropriate excitation energies, as discussed in Supplementary 
Information.

\section{Supplementary Information}

\begin{figure}[th!]
\begin{center}
\includegraphics[width=0.43\textwidth]{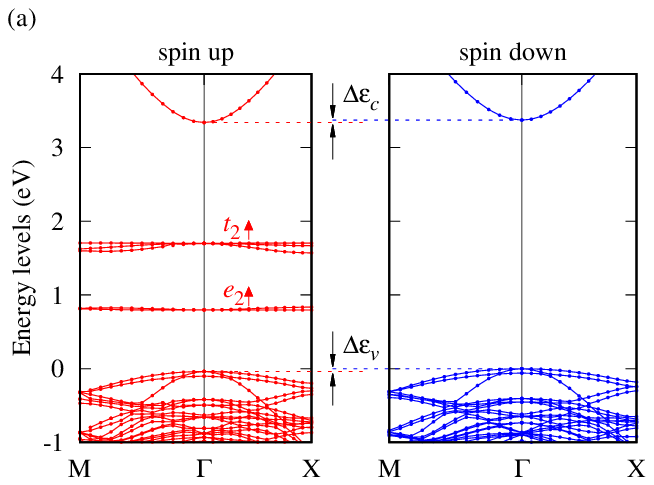}
\quad
\includegraphics[width=0.4\textwidth]{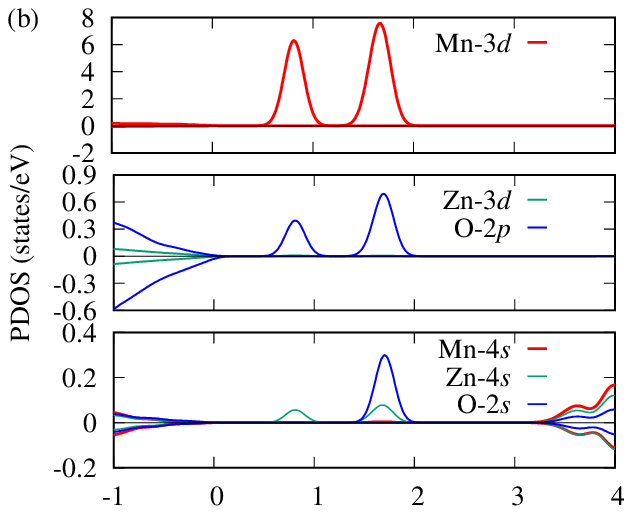}
\caption{\label{fig:Mn1}\small
(a) Spin-up and spin-down band structure, and (b) 
partial density of states of ZnO:Mn$^{2+}$ around the band gap. In (b),  
densities of one of the oxygen nearest-neighbour of Mn and of one of the 
next-nearest-neighbor zinc atom of Mn are shown, and thus their 
contributions to the gap states are relatively large. 
}
\end{center}

\begin{center}
\includegraphics[width=\textwidth]{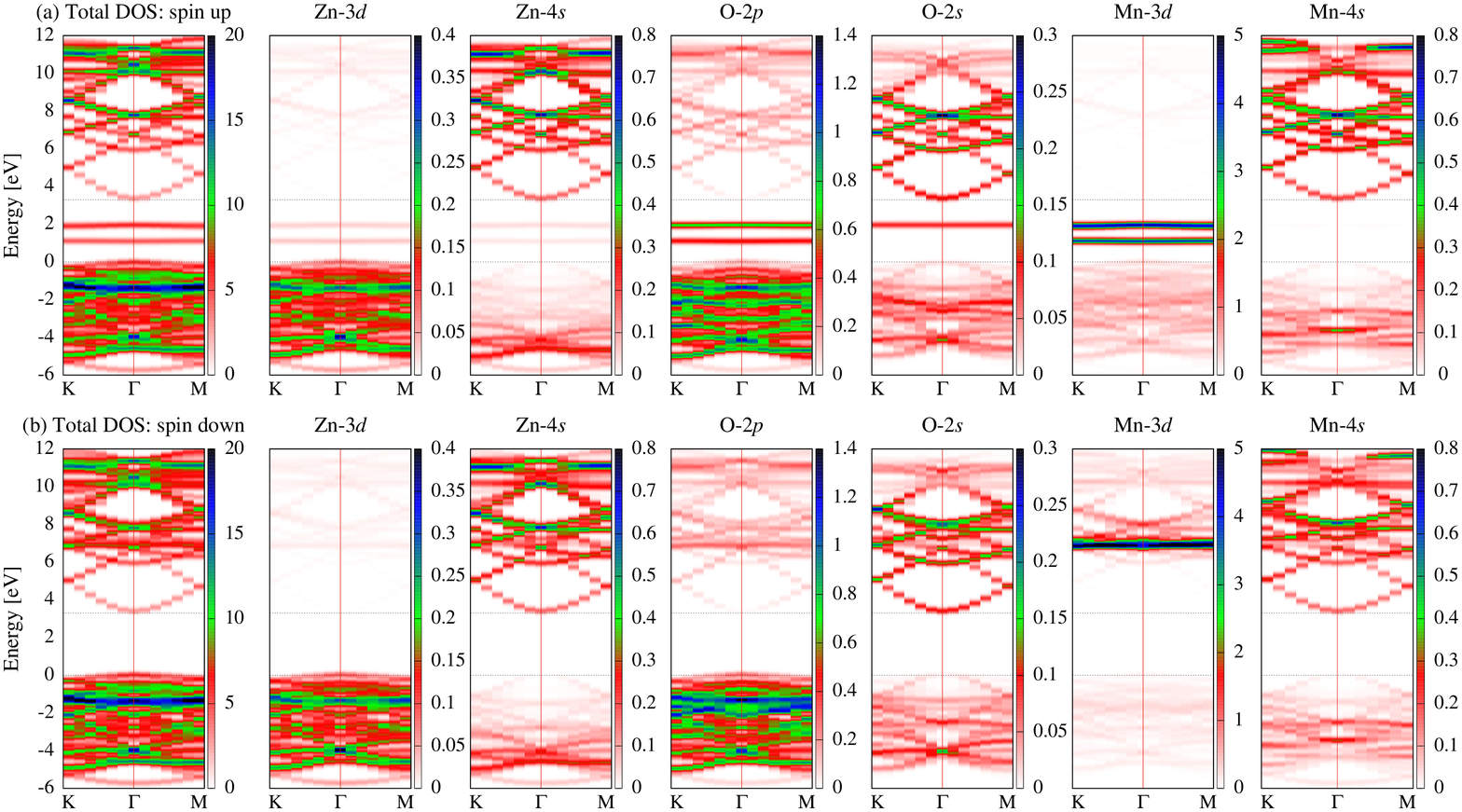}
\caption{\label{fig:Mn2}\small
(a) Spin-up and (b) spin-down part of k-resolved partial 
density of states of ZnO:Mn$^{2+}$. Note that the color-code scales on 
the consecutive figures are different. Oxygen and zinc atoms are the same 
as in Fig.~\ref{fig:Mn1}. 
}
\end{center}
\end{figure}

\subsection{Band structure}
In Fig.~\ref{fig:Mn1} (a), the energy bands of ZnO:Mn$^{2+}$ in the 
narrow energy window in the vicinity of both the CBM and the VBM are 
shown, together with their spin splittings. Figure~\ref{fig:Mn1} (b) 
shows the corresponding partial density of states (PDOS) for the spin-up 
and spin-down channels. Both figures clearly demonstrate that the CBM is 
composed from $s$(O) and $s$(Zn) orbitals with small contribution of 
$s$(Mn) (note that PDOSes of single O and Zn atoms are given).  
In turn, the VBM is composed mainly from $p$(O) orbitals with an addition 
of $d$(Zn). A small contribution of $d$(Mn) to spin-up valence band is an 
effect of the $p-d$ hybridization. The hybridization results also in 
the contribution from O ions to the Mn-induced gap levels (see the wave 
functions in the next section). 

Details of the orbital composition of states are shown in  
Fig.~\ref{fig:Mn2}, extending our previous analysis to the whole 
Brillouin Zone and a large energy window. In particular, the pronounced 
difference between the DOS of the $d$(Mn)-up  and the $d$(Mn)-down orbitals 
shows that $p-d$ hybridization is strongly spin-dependent. The $d$(Mn)-up 
contribute mainly to the two spin-up gap states, $t_2$ and $e_2$, which 
form almost dispersionless bands, but this contribution is non-vanishing 
for all valence bands that extend from 0 to -6 eV, and negligible for the 
conduction bands. On the other hand, the $d$(Mn)-down orbitals form a 
resonance above the CBM. Practically, they contribute only to the 
conduction bands. The added figures provide a supplementary insight into 
the $s,p-d$ hybridization and its consequences for $s,p-d$ coupling.

\subsection{Wave functions}
\begin{figure}[h]
\begin{center}
\includegraphics[width=0.49\textwidth]{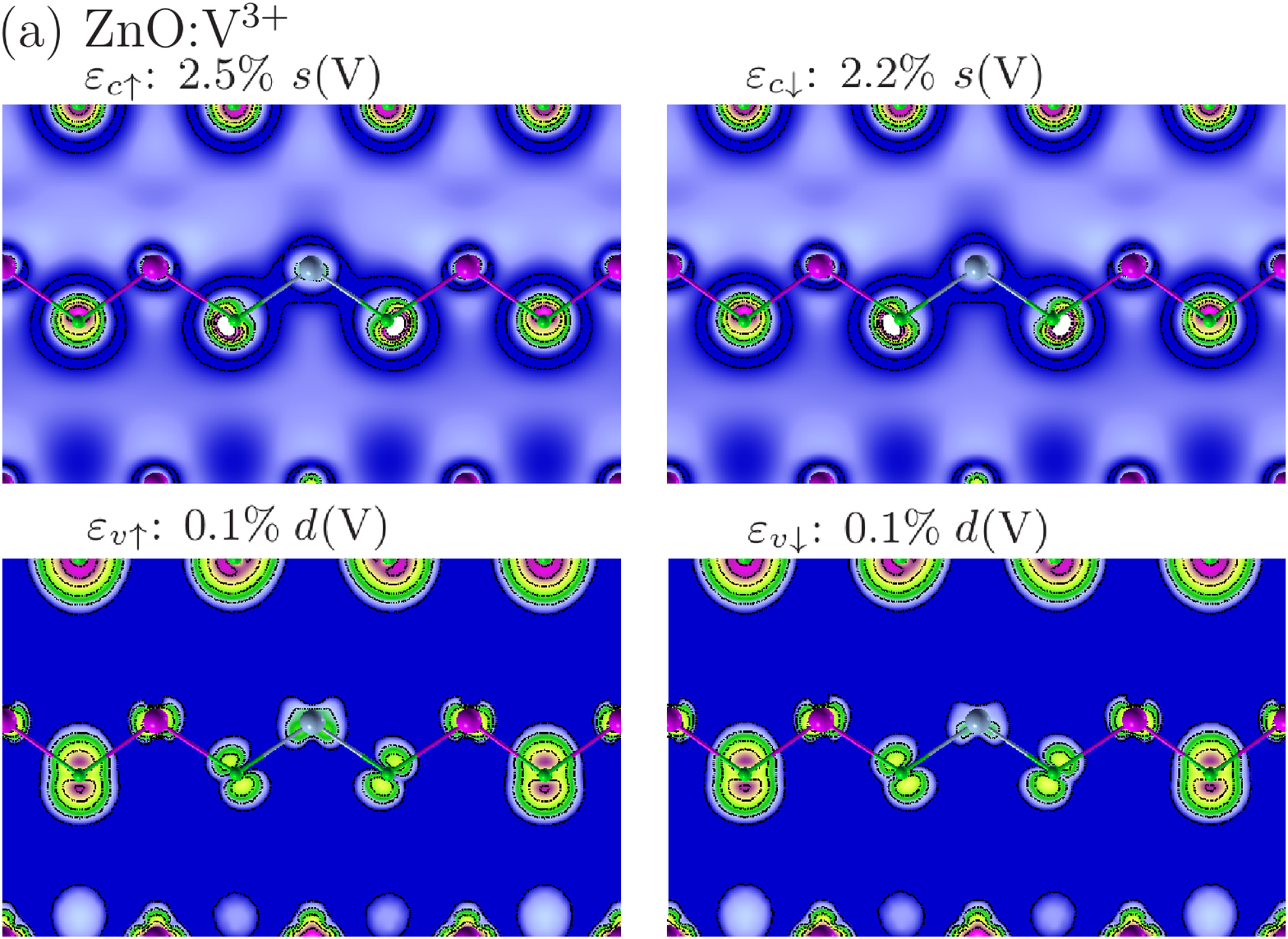}\quad
\includegraphics[width=0.49\textwidth]{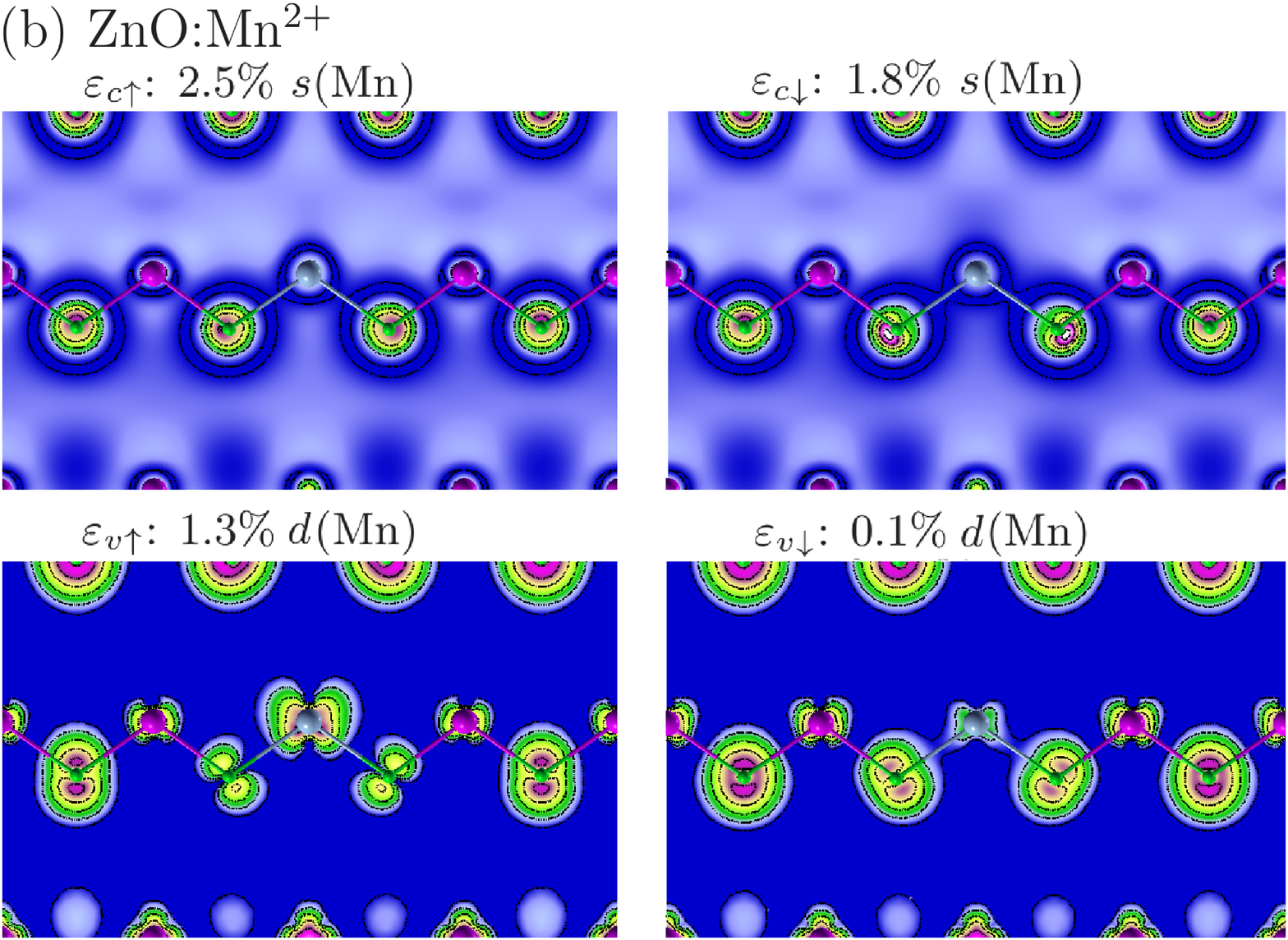}\\
\vspace{0.5cm}
\includegraphics[width=0.49\textwidth]{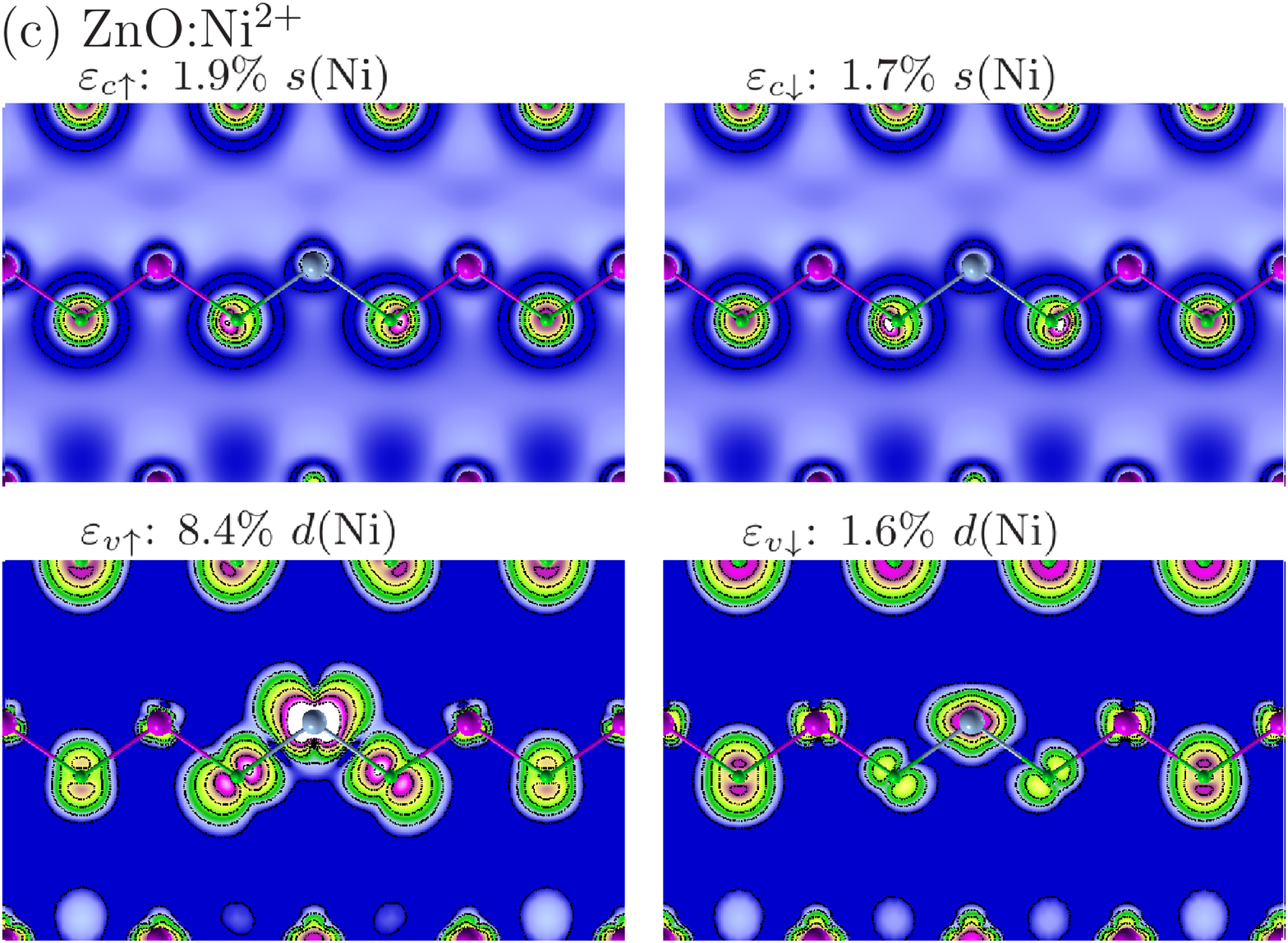}\quad
\includegraphics[width=0.49\textwidth]{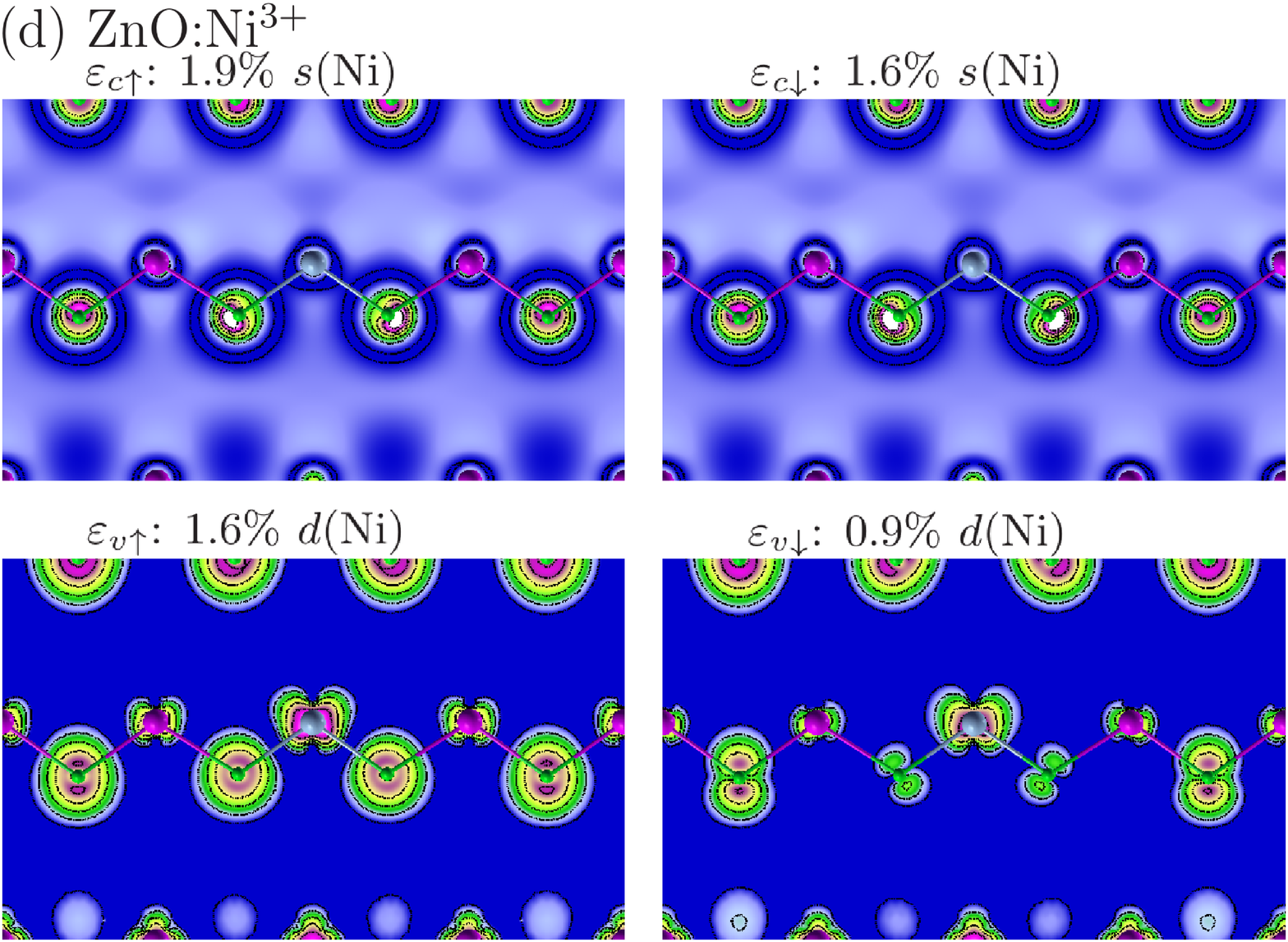}
\caption{\label{fig:wf_S}\small
The wave functions squared of the CBM (top panels) and 
the VBM (bottom panels) for 
(a) ZnO:V$^{3+}$, (b) ZnO:Mn$^{2+}$, (c) ZnO:Ni$^{2+}$ and (d) 
ZnO:Ni$^{3+}$. 
The contribution of TM states is given in each cases.  
}
\end{center}
\end{figure}

\begin{figure}[h]
\begin{center}
\includegraphics[width=0.75\textwidth]{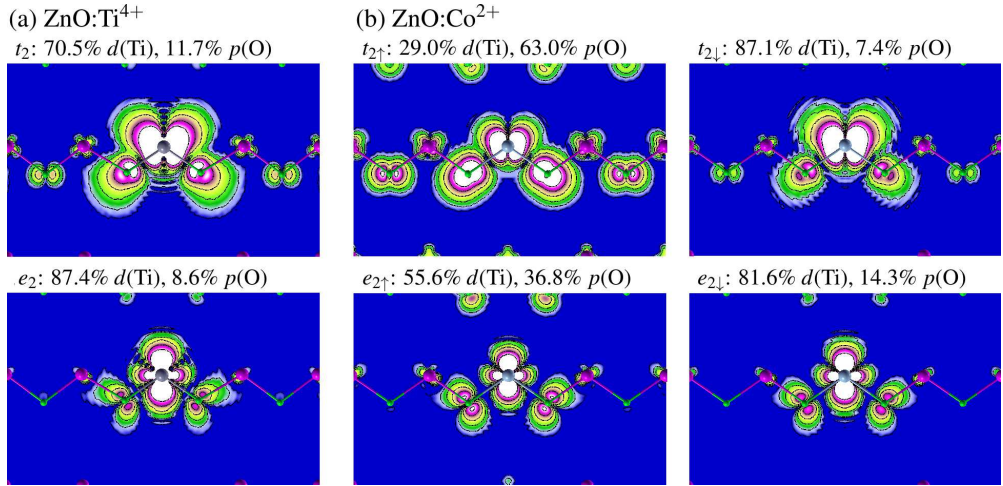}
\caption{\label{fig:TMwf}\small
The wave functions squared of the sum of $e_2$ (top 
panels) and the sum of $t_2$ (bottom panels) levels for (a) non-magnetic 
ZnO:Ti$^{4+}$, (b) magnetic ZnO:Co$^{2+}$. 
The contribution of $d$(TM) and $p$(O) states is given in each cases.  
}
\end{center}
\end{figure}

Figure~\ref{fig:wf_S} shows the wave functions of the CBM and VBM for V, 
Mn and Ni in ZnO. The CBM wave functions are similar for all TM ions, and 
both spin-up and spin-down partners can be treated as a slightly 
perturbed CBM of the pure ZnO. 
In contrast, the VBM states strongly depend on the TM ion and its charge 
state. The hybridization between the impurity $d$(TM) and the host $p$(O) 
orbitals increases as the energy difference between $t_{2\sigma}$(TM) and 
the VBM decreases. 
For V$^{3+}$ with 2 electrons in the $d$ shell, both  $t_{2\uparrow}$ and 
$t_{2\downarrow}$ triplets are empty and lie well above the CBM. 
Therefore, their contributions to the VBM are comparable and relatively 
small. The wave functions react locally to an impurity, but the response 
of the spin-up and -down functions is similar, leading to small 
$N_0\beta$s.
Next, Mn$^{2+}$ (with 5 $d$ electrons) is characterized by the fully 
occupied $t_{2\uparrow}$ level in the gap and the empty $t_{2\downarrow}$ 
above the CBM. 
The $p-d$ hybridization concerns mainly the spin-up channel, which 
determines the sign and strength of the exchange coupling. 
Regarding Ni$^{2+}$ with 8 $d$ electrons, the occupied $t_{2\uparrow}$ is 
very close to the VBM, giving a strong contribution to the $VBM\uparrow$. 
But the singlet derived from $t_{2\downarrow}$ is also in the gap, and 
its  considerable hybridization with the VBM effectively reduces 
$N_0\beta$.
A different situation takes place for Ni$^{3+}$ with 7 $d$ electrons.  It 
generates the fully occupied $t_{2\uparrow}$ level below the VBM, and  
hybridization between those states leads to the negative $N_0\beta$. 
Moreover, the empty $t_{2\downarrow}$ is in the gap, and its coupling 
with the $VBM\downarrow$ gives an  additional negative contribution to 
$N_0\beta$.

It is worth to mention that
hybridization results in the contribution of $p$(O) electrons to the TM-
induced levels as well. Because of symmetry, the $p-d$ 
mixing which originates from the VBM states concerns $t_{2}$ states, 
while the contribution of $p$(O) to $e_2$ states is an effect of mixing 
with states below the VBM. 
As it follows from Fig.~\ref{fig:TMwf}, the hybridization applies to 
both magnetic and non-magnetic impurities. However, in the case of the 
non-magnetic one, like Ti$^{4+}$, the is no difference between spin-up 
and spin-down levels. They are characterized by the same energy and the 
same contribution from $p$(O) and $d$(TM) states. 
Besides, because of Ti levels are much higher in energy than Co levels, 
the $p-d$ hybridization for Ti is weaker. In turn, for magnetic 
Co$^{2+}$, there is a large difference between $p-d$ hybridization for 
spin-up and spin-down electrons. 
Finally, one can note a pronounced 
hybridization-induced delocalization of the TM gap states, especially 
when compared with the relatively compact spin polarization $\Delta n$ 
shown in Fig.~\ref{fig:wf} of the paper.

\begin{figure}[th!]
\begin{center}
\includegraphics[angle=-90,width=0.5\textwidth]{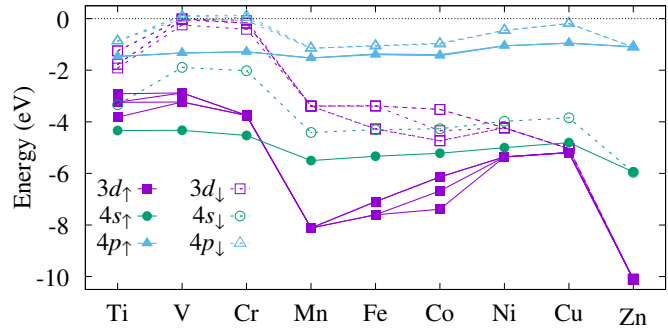}
\caption{\label{fig:TMatoms}\small
 Single particle levels of TM atoms for $q=0$. 
}
\end{center}
\end{figure}
\subsection{\label{TMatom}TM atoms}
Energies of isolated TM atoms are calculated using cubic supercells with 
the 20~\AA\ edge, sufficient to decouple adjacent atoms, and the same GGA 
pseudopotentials and cutoff energies as in the paper.

Calculated properties of the TM dopants in ZnO reflect those of 
individual TM atoms. Unfortunately, as it was pointed out already by, 
e.g., Janak~\cite{Janak} the DFT calculations encounter problems when 
applied to isolated TM atoms, because the self-consistent solutions 
giving the energy minimum are obtained for fractional occupations of both 
3$d$ and 4$s$ shells. 
Such configurations are not acceptable based on general arguments, and 
also they give somewhat distorted 3$d$ and 4$s$ energies. The issue was 
discussed for Fe in ZnO.~\cite{Fe} 
According to our calculations, Mn is the only atom for which the correct 
integer occupations are obtained, namely $d^5s^2$ for $q=0$ and $d^5s^1$ 
for $q=+1$ charge state. 
For the remaining atoms and for both $q=0$ and $q=+1$, the fractional 
occupations are found. 
Our $U$(TM)$=0$ results, shown in Fig.~\ref{fig:TMatoms}, are close to 
the  
data obtained by LSDA.~\cite{Kraisler, Shirley}  
A characteristic feature is the non-monotonic dependence of the $d$-shell 
energies on the atomic number. It explains the non-monotonic dependence 
of the $t_2$ and $e_2$ gap states found for 
the series Mn--Cu in ZnO. Eigenenergies of the singly ionized $q=+1$ 
atoms are 
lower by about 4--5~eV, but this feature persists. Only for $q=+2$ a 
monotonic dependence takes place.

\subsection{\label{sec:exc}$N_0\alpha$ and $N_0\beta$ from excitation energies}
The exchange constants of ZnO:TM discussed in the work were calculated 
directly from the spin splitting of the conduction and the valence bands, 
see Fig.~\ref{fig:Mn1}. On the other hand, experimental determination of 
$N_0\alpha$ and $N_0\beta$ often relies on magnetooptical experiments. 
The measured energies of excitonic transition can be directly compared 
with the calculated energies of the excited states of ZnO:TM, which 
provides an alternative to the approach based on the Kohn-Sham single 
particle levels used in the paper. To check the consistency of those two 
approaches, we calculated $\Delta \varepsilon_c$ as a difference in total 
energy of a supercell with one additional electron at the CBM, either on 
the spin-up or on the spin-down state. $\Delta \varepsilon_v$ is 
calculated comparing supercells with one spin-up or spin-down hole at the 
VBM. The TM charge state is ensured by fixing the occupation numbers of 
all single particle levels. The comparison was performed for Cr and Mn, 
for which all occupied dopant levels are well defined in the band gap. 
Difference in $N_0\alpha$ obtained by both methods is less than 0.01~eV, 
while that in $N_0\beta$ is less than 0.05~eV, so they can be treated as 
equivalent.

\section*{Acknowledgements}
The authors acknowledge the support from the Projects No. 
2016/21/D/ST3/03385, which are financed by Polish National Science Centre 
(NCN). Calculations were performed on ICM supercomputers of University of 
Warsaw (Grant No. GB77-15 and G16-11). We thank A. {\L}usakowski for the 
critical reading of the manuscript.

\bibliography{ZnObiblio2020}

\end{document}